\documentclass[apj]{emulateapj}
\pdfoutput=1
\usepackage{amsmath}
\usepackage{mathrsfs}

\newcommand{\thco}{$^{13}$CO}
\newcommand{\ceo}{C$^{18}$O}
\newcommand{\nthp}{N$_2$H$^{+}$}
\newcommand{\htdp}{H$_2$D$^{+}$}
\newcommand{\hcop}{HCO$^{+}$}
\newcommand{\hii}{H\,{\sc ii}}
\newcommand{\kms}{km\,s$^{-1}$}
\newcommand{\msun}{$M_{\odot}$}
\newcommand{\lsun}{$L_{\odot}$}

\newcommand{\eg}{{\it e.g.}}

\newcommand{\cmt}{cm$^{-3}$}
\newcommand{\mcloudb}{{\it MSX}\,G028.53$-$00.25}
\newcommand{\mclouda}{{\it MSX}\,G030.88+00.13}
\newcommand{\spitzer}{{\em Spitzer}}
\newcommand{\lvsm}{$L_{\rm bol}$--$M_{\rm env}$}
\begin{document}
\slugcomment{Accepted by the Astrophysical Journal}
\shorttitle{Two Massive, Low-Luminosity Cores}
\shortauthors{Swift}
\title{Two Massive, Low-Luminosity Cores Toward Infrared Dark Clouds}
\author{Jonathan J. Swift}
\email{js@ifa.hawaii.edu}
\affil{Institute for Astronomy, 2680 Woodlawn Dr., Honolulu, HI
96822-1897}

\begin{abstract}
This article presents high-resolution interferometric mosaics in the 
850\,\micron\ waveband of two massive, quiescent infrared dark
clouds. The two clouds were chosen based on their likelihood to
represent environments preceding the formation of massive stars. The
brightest compact sources detected in each cloud have masses $\approx
110$\,\msun\ and $\approx 60$\,\msun\ with radii $< 0.1$\,pc, implying
mean densities of $\langle n\rangle\approx 10^6$\,cm$^{-3}$ and
$\langle N\rangle\approx 1$\,g\,cm$^{-2}$. Supplementary data show
these cores to be cold and inactive. Low upper limits to their
bolometric luminosities and temperatures place them at a very early
stage of evolution while current models of massive star formation
suggest they have the potential to form massive stars.
\end{abstract}


\section{Introduction} \label{Intro}
It is fairly well established that low-mass stars form
from the gravitational collapse of dense condensations, or
“cores”\footnote{Following the convention presented in \cite{wil00},
  ``cores'' describe dense 
  condensations of molecular gas with $M \lesssim 100$\,\msun\ that
  will form individual stars or small stellar systems, while
  ``clumps'' describe dense cloud structures with mass
  $\gtrsim 1000$\,\msun\ that are more likely to form clusters of
  stars}, 
within molecular clouds \citep{mye83b,bei86,mot98}, and there are many
examples of cores in nearby star-forming regions ($\lesssim 300$\,pc)
that appear to represent the direct progenitors of low-mass stars, at
least in a statistical sense \citep[\eg,][]{alv07}. However,
surveys of distant regions where massive stars form suffer from poor 
physical resolution, blending together regions with sizes comparable
to clusters of stars \citep[{\eg},][]{shi03,eva08}. 
High-resolution interferometric observations can resolve these massive
clumps into substructure more closely related to individual stars or
stellar systems \citep[\eg,][]{beu07a,pil06b,mol02}. But in virtually
all cases, these high-resolution studies have targeted regions with
indications for young massive stars such as bright mid-infrared
emission, masers, outflow, or compact \hii. 

This article presents two interferometric mosaics of infrared dark
clouds \citep[IRDCs;][]{sim06a,per96,ega98} selected to lack
massive protostars yet likely to be pre-cluster clouds. Several dense
cores are detected within each IRDC, but it is the 
most massive cores in each cloud are the focus of this
article. Following a description of our observations in \S\,\ref{Obs},
the general properties of the IRDCs are described in \S\,\ref{Cloud}
using new single dish data together with published data. The
interferometric mosaics are then considered in \S\,\ref{SMosaic} where
the physical properties of the massive cores are presented and
discussed. 

\section{Observations and Data Reduction} \label{Obs}
The infrared dark clouds \mclouda\ and \mcloudb\ were chosen from the 
catalog of \cite{sim06b} using \spitzer\ Galactic plane survey
data \citep{ben03,car09}, and SCUBA archival data \citep{dif08} to be
massive, dense, and lack significant 24\,\micron\ emission. These
attributes were desired to maximize the probability that the clouds are
in a state preceding the formation of massive stars. The observations
are summarized in Table~\ref{obstable}.  
\renewcommand{\arraystretch}{0.9}
\begin{deluxetable*}{lccccc}
\tablewidth{0in}
\tablecaption{Summary of Observations}
\tablehead{\colhead{Date} &
\colhead{Waveband} &
\colhead{Conditions} &
\colhead{Target\tablenotemark{a}} &
\colhead{Calibrators} &
\colhead{Obs. Type} }
\startdata
\multicolumn{6}{c}{The Submillimeter Array} \\ \cline{1-6}
 2007Jun24 & 350.9\,GHz\tablenotemark{b}   & $\tau_{225} \approx 0.08$\tablenotemark{c} & B   & 1751+096/1743-038 & 1   \\
 2007Jul01 & 351.0\,GHz\tablenotemark{b}   & $\tau_{225} \approx 0.08$\tablenotemark{c} & B   & 1751+096/1743-038 & 1   \\
 2007Jul07 & 350.9\,GHz\tablenotemark{b}   & $\tau_{225} \approx 0.08$\tablenotemark{c} & A   & 1751+096/1743-038 & 1   \\
 2007Oct19 & 350.9\,GHz\tablenotemark{b}   & $\tau_{225} \approx 0.08$\tablenotemark{c} & A   & 1751+096/1743-038 & 1   \\
 2008Jun02 & 340.5\,GHz\tablenotemark{b}   & $\tau_{225} \approx 0.15$\tablenotemark{c} & A,B & 1751+096/1911-201 & 1,2 \\
 2008Jun15 & 340.5\,GHz\tablenotemark{b}   & $\tau_{225} \approx 0.11$\tablenotemark{c} & A,B & 1751+096/1911-201 & 1,2 \\
\cutinhead{James Clerk Maxwell Telescope}
 2007Jul17 & 334.8\,GHz/350.7\,GHz\tablenotemark{b} & $\tau_{225} \approx 0.10$\tablenotemark{c} & A,B & V437Sct/16293-2422/G45.1    & 3  \\
 2007Jul18 & 350.7\,GHz/367.3\,GHz\tablenotemark{b} & $\tau_{225} \approx 0.06$\tablenotemark{c} & B   & V437Sct/16293-2422/HD235858 & 3,4 \\
 2007Jul19 & 334.8\,GHz/350.7\,GHz/367.3\,GHz\tablenotemark{b} & $\tau_{225} \approx 0.06$\tablenotemark{c} & A,B & V437Sct/16293-2422/G34.2    & 3,4 \\
 2007Jul20 & 334.8\,GHz/350.7\,GHz/367.3\,GHz\tablenotemark{b} & $\tau_{225} \approx 0.07$\tablenotemark{c} & A,B & V437Sct/16293-2422/G34.2    & 3,4 \\
 2008Aug06 & 367.3\,GHz\tablenotemark{b}                       & $\tau_{225} \approx 0.05$\tablenotemark{c} & A & WAql    & 4 \\
 2008Nov12 & 367.3\,GHz\tablenotemark{b}                       & $\tau_{225} \approx 0.07$\tablenotemark{c} & A & HD179821/NMLCyg & 4 \\
\cutinhead{Canda France Hawaii Telescope}
 2008Aug10 & $J$, $H$, $K_s$ & $\theta_s \approx 0.6^{\prime\prime}$\tablenotemark{d} & A,B & 2MASS & 5 \\
 2008Aug16 & $J$, $H$, $K_s$ & $\theta_s \approx 0.7^{\prime\prime}$\tablenotemark{d} & A,B & 2MASS & 5 
\enddata 
\tablecomments{(1) Mosaic observations, (2) Targeted (single pointing)
  observations, (3) raster scan mapping, (4) jiggle chop observations
  (5) direct imaging.}
\tablenotetext{a}{Target A corresponds to \mclouda, and target B
  corresponds to \mcloudb.}
\tablenotetext{b}{\,Local oscillator frequency.}
\tablenotetext{c}{Atmospheric opacity at 225\,GHz.}
\tablenotetext{d}{Full width at half maximum of the point
  spread function.}
\label{obstable}
\end{deluxetable*}

The target IRDCs were observed with the Submillimeter
Array\footnote{The Submillimeter Array is a joint project between the
  Smithsonian Astrophysical Observatory and the Academia Sinica
  Institute of Astronomy and Astrophysics and is funded by the
  Smithsonian Institution and the Academia Sinica.} 
(SMA) in its compact configuration using the 345\,GHz
receivers. Mosaic observations were designed to cover the
brightest SCUBA emission in each IRDC. These observations preserved
the full 4\,GHz bandwidth and specified both high and low resolution
correlator chunks with $0.17$ and $2.8$\,\kms\ channel widths
respectively. The data were calibrated using the MIR reduction package
\footnote{{\tt
    http://www.cfa.harvard.edu/$\sim$cqi/mircook.html}}. Visibility
phases were corrected using quasar observations conducted in 25
minute intervals. The bandpass was calibrated with observations of the
quasar 3C279, and Uranus was used to tie down the flux scale to an
accuracy of $\sim 15$\%. The visibility data were then output into
MIRIAD\footnote{{\tt http://bima.astro.umd.edu/miriad}} format and
inverted into the image domain. The mean rms level across the mosaic
maps for \mclouda\ and \mcloudb\ are 5.4\,mJy and 4.6\,mJy for
synthesized beam sizes of $1.9^{\prime\prime}\times1.8^{\prime\prime}$
and $2.0^{\prime\prime}\times1.1^{\prime\prime}$ oriented at position
angles of $66^\circ$ and  $68^\circ$, respectively.

The James Clerk Maxwell Telescope\footnote{The James Clerk Maxwell 
  Telescope is operated by The Joint Astronomy Centre on behalf of the
  Science and Technology Facilities Council of the United Kingdom, the
  Netherlands Organisation for Scientific Research, and the National
  Research Council of Canada.}(JCMT) equipped with the HARP-B receiver
array and ACSIS backend \citep{den00} was used to observe our target
IRDCs over several nights. Spectral lines of CO, \thco, \ceo, \nthp,
and \htdp\ were placed in spectral windows of 250\,MHz width and
61\,kHz resolution. Flux levels are expected to be accurate to 15\%
and the pointing accuracy of the final maps are estimated to be better
than $3.5^{\prime\prime}$. The Starlink software suite was used to grid and
output the data into FITS format and IDL was used to perform final
calibrations and coadding. A main beam efficiency $\eta_{\rm MB} =
0.7$ is used to convert corrected antenna temperatures to main beam
temperatures.

The Canada-France Hawaii Telescope\footnote{The Canada-France-Hawaii
  Telescope (CFHT) is 
  operated by the National Research Council of Canada, the Institut
  National des Sciences de l'Univers of the Centre National de la
  Recherche Scientifique of France, and the University of Hawaii.}
(CFHT) was used to observe our target clouds using the WIRCam infrared
detector \citep{pug04}. The observations were carried out in queue
mode over the course of two photometric nights. The pre-processed data
(de-biased, flat-fielded and sky-subtracted by the CFHT pipeline) were
downloaded and further reduced using the TERAPIX\footnote{{\tt
http://terapix.iap.fr}} software suite. The images were registered,
combined by weighted mean, and tied to the 2MASS\footnote{The Two
  Micron All Sky Survey is a joint project of the University
  of Massachusetts and the Infrared Processing and Analysis 
  Center/California Institute of Technology, funded by the National
  Aeronautics and Space Administration and the National Science
  Foundation.} point source catalog flux scale to an accuracy of
better than 0.02 magnitudes. 

\section{The IRDC environments} \label{Cloud}
\begin{figure*}[!t]
\centering
\includegraphics[angle=0,width=6in]{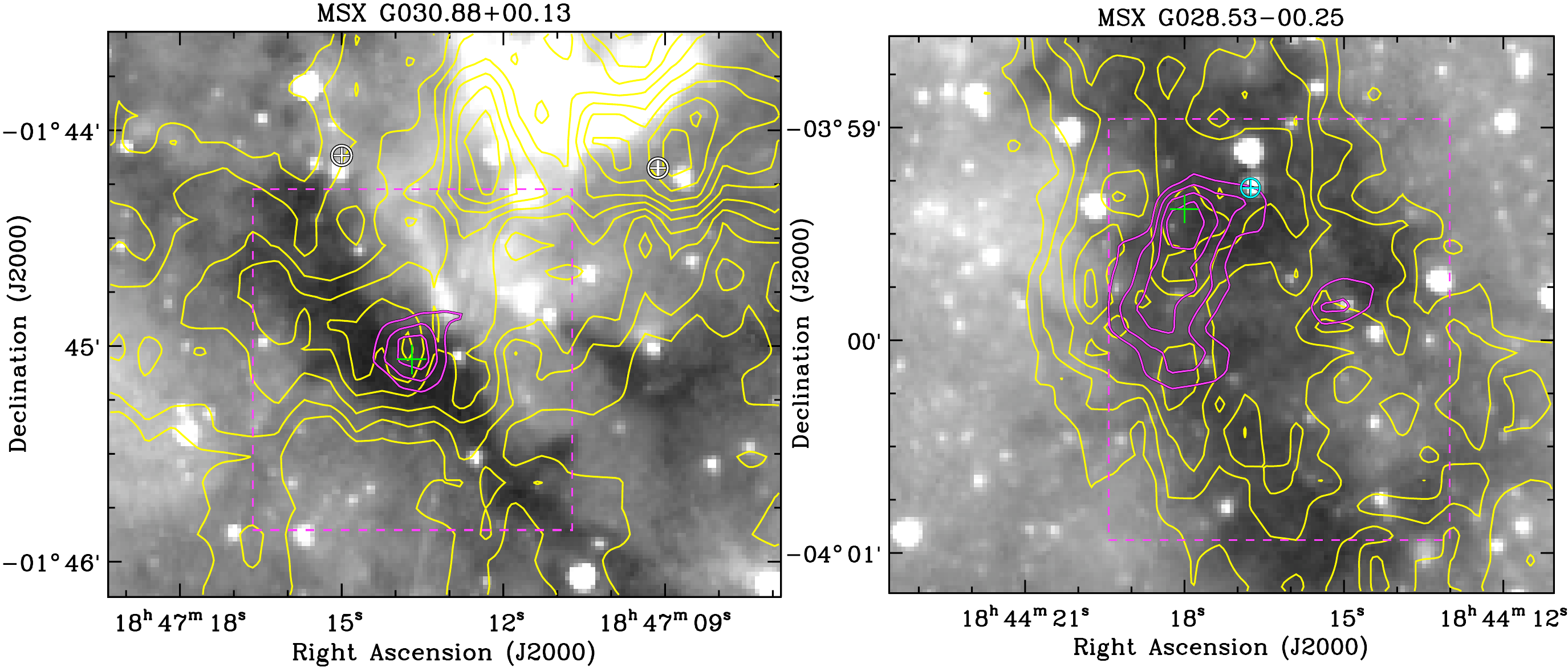}
\caption{Gray scale images of $5.8 + 8.0$\,\micron\ emission toward
  two infrared dark clouds. Contours of \thco$(3-2)$ emission are
  shown in yellow and contours of \nthp$(4-3)$ are shown in
  magenta. The circled crosses denote the positions of methanol ({\em
    white}) and water ({\em cyan}) masers. The positions of the
  brightest compact cores detected by the SMA (see \S\,\ref{SCores})
  are shown with the 
  green cross.
 \label{IRDCs}}
\end{figure*}

Figures~\ref{IRDCs}a and \ref{IRDCs}b show images from the
\spitzer\ GLIMPSE survey in which the mid-infrared extinction features
that define the clouds can be clearly seen. Contours of velocity
integrated \thco$(3-2)$ overlay the images in yellow. \mclouda\ has 2
velocity components along the line-of-sight with centroids at 95 and
107\,\kms\ with respect to the local standard of rest. The higher
velocity component shown in Figure~\ref{IRDCs}a at \thco\ emission
levels of $(5 + 3\,n)$\,K\,\kms\ $(n = 0,1,2...)$ is widespread across
the region suggesting that the IRDC is part of a larger complex
including the bright infrared cluster to the northwest. \mcloudb\ has
a single velocity component in \thco\ at 87\,\kms\ shown at contour
levels of $(6+2\,n)$\,K\,\kms\ seen to be spatially confined to
the region of mid-infrared extinction. The presence of high volume
density gas in these clouds is confirmed with detections of
\nthp$(4-3)$ shown as magenta contours at levels of $(1 +
0.5\,n)$\,K\,\kms\ and $(1 + 0.4\,n)$\,K\,\kms\ for
Figures~\ref{IRDCs}a and \ref{IRDCs}b, respectively.

The velocity centroid of \mclouda\ places it at the tangent point of
Galactic rotation in this direction. Therefore a distance of
$7.2$\,kpc can be derived using the simple geometric relationship $d =
R_0\cos(l)\cos(b)$, where $R_0$ is taken to be $8.4$\,kpc
\citep{ghe08,rei09}. The distance to \mcloudb\ is taken to be 5.4\,kpc
\citep{rat06}. The errors on these kinematic distances may be 15\% or
more due to non-circular motions in the Galaxy
\citep[{\eg},][]{rom09}. The masses of the IRDCs are estimated to be
several thousand solar masses based on CO isotopologue emission
assuming 15\,K gas in local thermodynamic equilibrium, SCUBA
850\,\micron\ continuum emission, and virial equilibrium.

There is no $\lambda 20$\,cm or $\lambda 6$\,cm emission detected
toward either IRDC \citep{hel06,whi05}. A total $\lambda 20$\,cm flux
of 250\,mJy seen toward the IR bright cluster in Figure~\ref{IRDCs}a
suggests the presence of a B0 star \citep{giv05b}. Two class II
methanol masers are seen in the field of \mclouda\ with velocities in
rough agreement with the cloud \citep{pes05}. No class II masers are
seen in the \mcloudb\ field. However, one water maser has been
detected toward an embedded source in this cloud \citep{wan06}. Fits
to the spectral energy distributions of MIPS 24\,\micron\ sources
within the IRDCs suggest they are not massive stars
\citep{rob06,rob07}. Although given the large column depths observed
toward the IRDCs and their distances, it is possible that more
protostars are embedded in the cloud than are detected in the
\spitzer\ data. We estimate that less than $\sim 100$ stars are
embedded within the IRDC based on extrapolations from mid-infrared
emission from nearby, intermediate-mass star forming regions
\citep{pad08,reb07} and a Salpeter stellar mass function with a
lognormal turnover \citep{sal55,cha03}.

\section{High-Resolution Sub-millimeter Wave Imaging} \label{SMosaic} 
\begin{figure*}[!t]
\centering
\includegraphics[angle=0,width=6in]{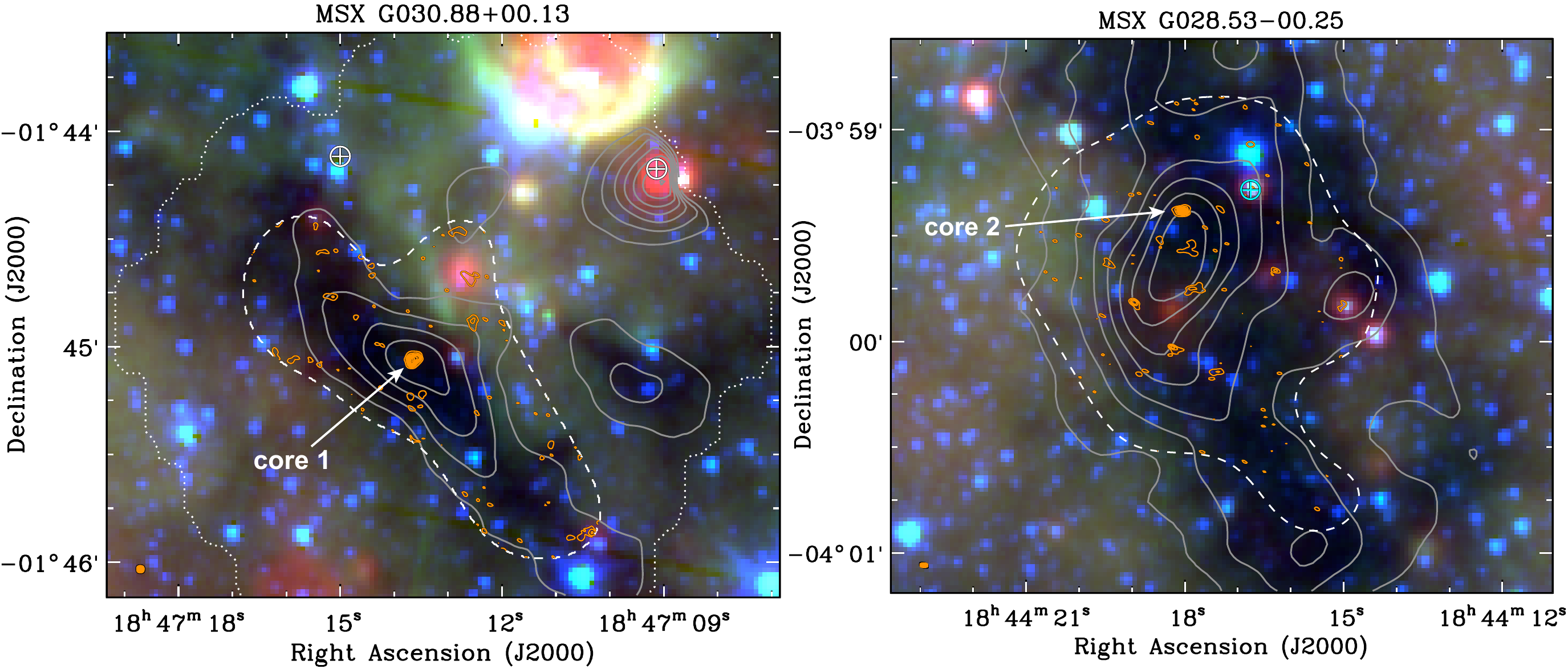}
\caption{\spitzer\ composite images (red = MIPS 24\,\micron, green =
  IRAC 8.0\,\micron\ + 5.8\,\micron, blue = IRAC 4.5\,\micron\ +
  3.6\,\micron) overlaid with contours of sub-millimeter continuum
  emission from the SMA mosaic data at levels of $3\sigma\,n$ ({\em
    orange}) and SCUBA archival data shown at levels of $250\,n$\,mJy
  ({\em gray}) with $n = 1,2,3,..$. The coverage of the SMA and SCUBA
  data are outlined with a dashed and dotted contour, respectively.}
\label{SpitzerSMA}
\end{figure*}

Figures~\ref{SpitzerSMA}a and \ref{SpitzerSMA}b show the contours of
the SMA mosaics overlaid on \spitzer\ mid-infrared composite
images. The brightest sources in each mosaic
dominate the compact 
sub-millimeter wave emission toward the IRDCs. These massive cores
lie near the near the peak of the SCUBA emission and are isolated with
respect to other compact sources. They are also isolated in mass with
the next most massive cores in each cloud being 
more than a factor of 3 less massive. Table~\ref{cores} displays the
physical characteristics of the two compact submillimeter sources.

\renewcommand{\tabcolsep}{2mm}
\begin{deluxetable*}{lccccccccccc}
\tablewidth{0in}
\tablecaption{Massive, Low-Luminosity Cores}
\tablehead{ \colhead{} &
\colhead{R.A.} &
\colhead{Decl.} &
\colhead{$F_{\rm peak}$} &
\colhead{$F_{\rm int}$} &
\colhead{Mass\tablenotemark{a}} &
\colhead{$r$\tablenotemark{b}} &
\colhead{$\langle n\rangle$} &
\colhead{$\langle N\rangle$} &
\colhead{$T_{\rm env}$} &
\colhead{$T_{\rm bol}$} &
\colhead{$L_{\rm bol}$} \\
\colhead{Name} &
\colhead{$(J2000)$} &
\colhead{$(J2000)$} &
\colhead{(mJy/bm)} &
\colhead{(mJy)} &
\colhead{(\msun)} &
\colhead{(pc)} &
\colhead{($10^6$\,\cmt)} &
\colhead{(g\,cm$^{-2}$)} &
\colhead{(K)} &
\colhead{(K)} &
\colhead{($L_\odot$)} }
\startdata
 Core 1\dotfill &  18:47:13.7 &  -01:45:03.7 &  119   & 255 & 110 & 0.082 & 0.82 & 1.1 & $<19$ & $<31$ & $<460$ \\
 Core 2\dotfill &  18:44:18.0 &  -03:59:23.0 &  118   & 224 &  60 & 0.056 &  1.5 & 1.3 & $<18$ & $<30$ & $<170$
\enddata
\tablenotetext{a}{Masses estimated using a dust temperature of
  15\,K and an opacity $\kappa = 0.019$\,cm$^2$\,g$^{-1}$ derived from
  \cite[][Table 1]{oss94}.} 
\tablenotetext{b}{Radii are derived by deconvolving the synthesized
  beam from the effective radius, $r = \sqrt{R_{\rm eff}^2 - 
  R_{\rm beam}^2}$, where  $R_{\rm eff} = \sqrt{A/\pi}$ and $A$ is the 
  area contained within a 3\,$\sigma$ contour.} 
\label{cores}
\end{deluxetable*}

\subsection{Two Massive, Low-Luminosity Cores} \label{SCores} 
The derived masses of the two cores are $\approx 110$\,\msun\ and
$\approx 60$\,\msun\ with densities of $\langle n \rangle \approx
10^6$\,\cmt\ and $\langle N \rangle \approx 1$\,g\,cm$^{-2}$. Their
free fall, or dynamical timescales are thus a few $10^4$\,yrs implying
that they are most likely short lived structures. Both cores appear
marginally resolved, but are detected in the longest baseline data
suggesting that perhaps there are resolved and unresolved components.

\begin{figure}[!b]
\centering
\includegraphics[angle=0,width=3.3in]{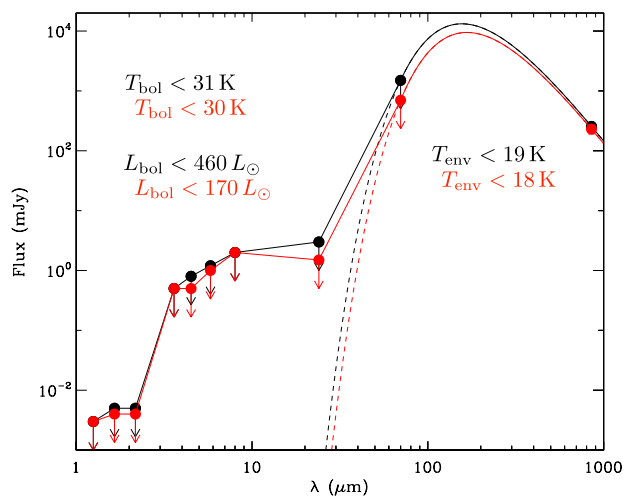}
\caption{Spectral energy distributions of core 1 ({\em black}) and
  core 2 ({\em red}). The detections at 853\,\micron\ have
  error bars smaller than the symbol size, while
  upper limits at wavelengths from 1.25 to 70\,\micron\ are
  shown. \label{maxtemp}}   
\end{figure}

Figure~\ref{maxtemp} shows the spectral energy distributions of the
cores consisting of upper limits from 1.25\,\micron\ to 70\,\micron\ and 
853\,\micron\ fluxes from Table~\ref{cores}. A best fit gray body curve
with $\tau = (\nu/\nu_{\rm c})^2$, and $\nu_{\rm c} = 6$\,THz
\citep{war02} limits the envelope
temperatures while summation under the solid curves
of Figure~\ref{maxtemp} provide upper limits to the bolometric
luminosities and temperatures \citep{mye93}.

High values of $M_{\rm env}/L_{\rm bol}^{0.6} \approx 2.8$ in solar
units and low values for $T_{\rm bol} \lesssim 30$\,K for 
these cores are signs of extreme youth. Class 0 protostars, the
earliest SED class, are marked by $M_{\rm env}/L_{\rm bol}^{0.6} >
0.4$ \citep{bon96,and00} and $T_{\rm bol} \lesssim
70$\,K \citep{che95}. Given the large distances to these cores and the
photometric sensitivity, the existence of a low luminosity sources
deeply embedded within the cores cannot be ruled out. However, it is
clear that no massive protostars exist in these cores.

Across the SMA bandpass, the cores are only detected in CO$(3-2)$ and
\hcop$(4-3)$. The CO$(3-2)$ is strongly filtered by the interferometer
making a useful interpretation of the emission from this low 
density gas tracer difficult. However, no clear indication of outflow is
seen toward either core in the single dish or interferometric
data further indicating their extreme youth. Upper limits
to SO$(8_9-7_8)$ and CH$_3$OH$(13_{7,7}-12_{7,6})$ emission highlight
the difference between these cores and the spectral signatures of hot
cores such as Orion KL \citep{sch97}.

\begin{figure}
\centering
\includegraphics[angle=0,width=3.3in]{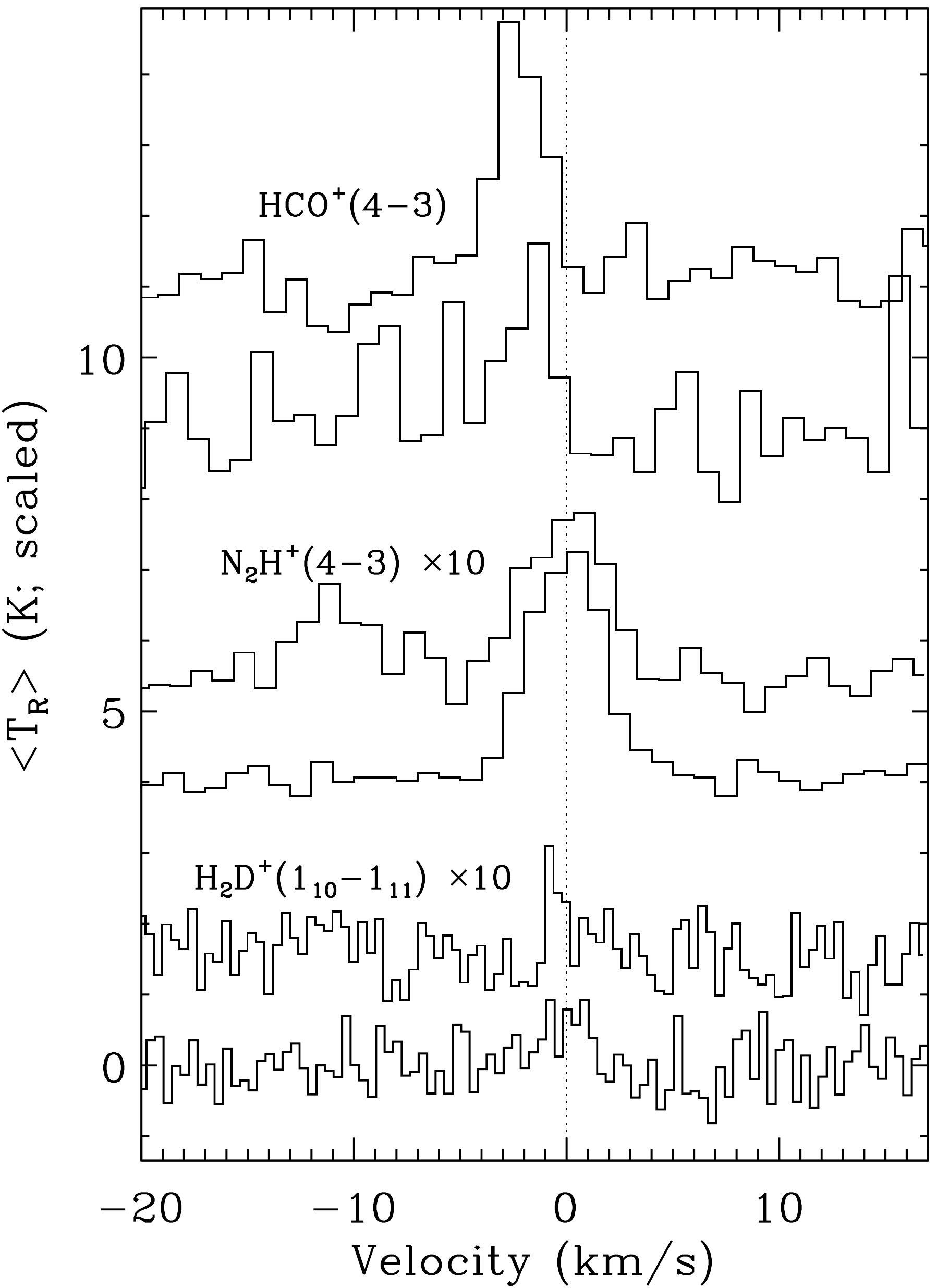}
\caption{Average spectra of high-dipole moment molecules toward core 1 (top)
  and core 2 (bottom) in molecular transitions tracing high volume
  density gas. The \nthp\ and \htdp\ spectra are scaled by a
  factor of 10 for the plot. Velocities are with respect to the
  centroid of \nthp\ emission for each core measured from a Gaussian
  fit to the line profiles. \label{dense_specs}} 
\end{figure}

Figure~\ref{dense_specs} shows the composite spectra of molecular
transitions tracing dense gas toward the two cores with the velocity
scale shifted relative to the centroid velocity of \nthp$(4-3)$
emission measured to be 107.2\,\kms\ and 86.8\,\kms\ for core 1 and 2,
respectively. Core 1 shows an \hcop\ emission feature that lies 
blueward of the systemic velocity. This may be an indication of
\hcop\ self-absorption, but there is also the possibility that the
emission in the central channels is widespread and filtered out by the 
interferometer. Core 2 shows a weak feature in the \hcop\ spectral
window blueward of its systemic velocity that may be a sign of a
similar spectral feature to core 1.  

Also shown in Figure~\ref{dense_specs} are spectra of
ortho-\htdp$(1_{1,0} - 1_{1,1})$ toward the cores. Core 1 shows an
emission feature with 4.3\,$\sigma$ significance centered at
106.6\,\kms\ with a full width of  $0.9\pm0.3$\,\kms. Core 2 shows a
broader feature with $2.0\pm0.7$\,\kms\ width at a 3.1\,$\sigma$
significance level. This molecular species is thought to trace cold,
dense, and chemically evolved gas in the deep interior of pre- or
proto-stellar cores \citep{wal04,vas06,cas08}. Following
\cite{cas08}, the integrated line fluxes of 0.13\,K\,\kms\ and
0.17\,K\,\kms\ for core 1 and 2 respectively translate to column
densities of $N($\htdp$) \approx 10^{11}$\,cm$^{-2}$ using a kinetic
temperature $T_{\rm k} = 15$\,K and a critical density $n_{\rm cr}
=10^6$\,\cmt. The fractional abundances of ortho-\htdp\ are
$N($\htdp$)/N($H$_2) \approx 3$--$5\times 10^{-13}$. Varying $T_{\rm k}$
from 10 to 15\,K and $N_{\rm cr}$ from $10^5$ to $10^6$\,\cmt\ changes
these values by a factor of about 2. These numbers are significantly
smaller than the values found for low-mass pre- and proto-stellar
cores \citep{vas06,cas08} as well as in Orion B \citep{har06}. This
could be due to an intrinsic dearth of \htdp. However, given the poor
physical resolution of our single dish observations and the propensity
for \htdp\ to trace the innermost regions of cores, it may be that
this discrepancy is due in large part to beam dilution.

\subsection{Discussion} \label{Discussion}
The signposts of massive star formation toward \mclouda\ lend credence 
to the idea that there might exist regions within the cloud where
massive stars will, but have not yet, formed. The characteristics of
core 1 make it a good candidate for one such region. While
\mcloudb\ currently shows no signs of high-mass star formation, core 2
shares similar traits to core 1 as being a potential massive star
precursor. 

\begin{figure*}[!t]
\centering
\includegraphics[angle=0,width=5in]{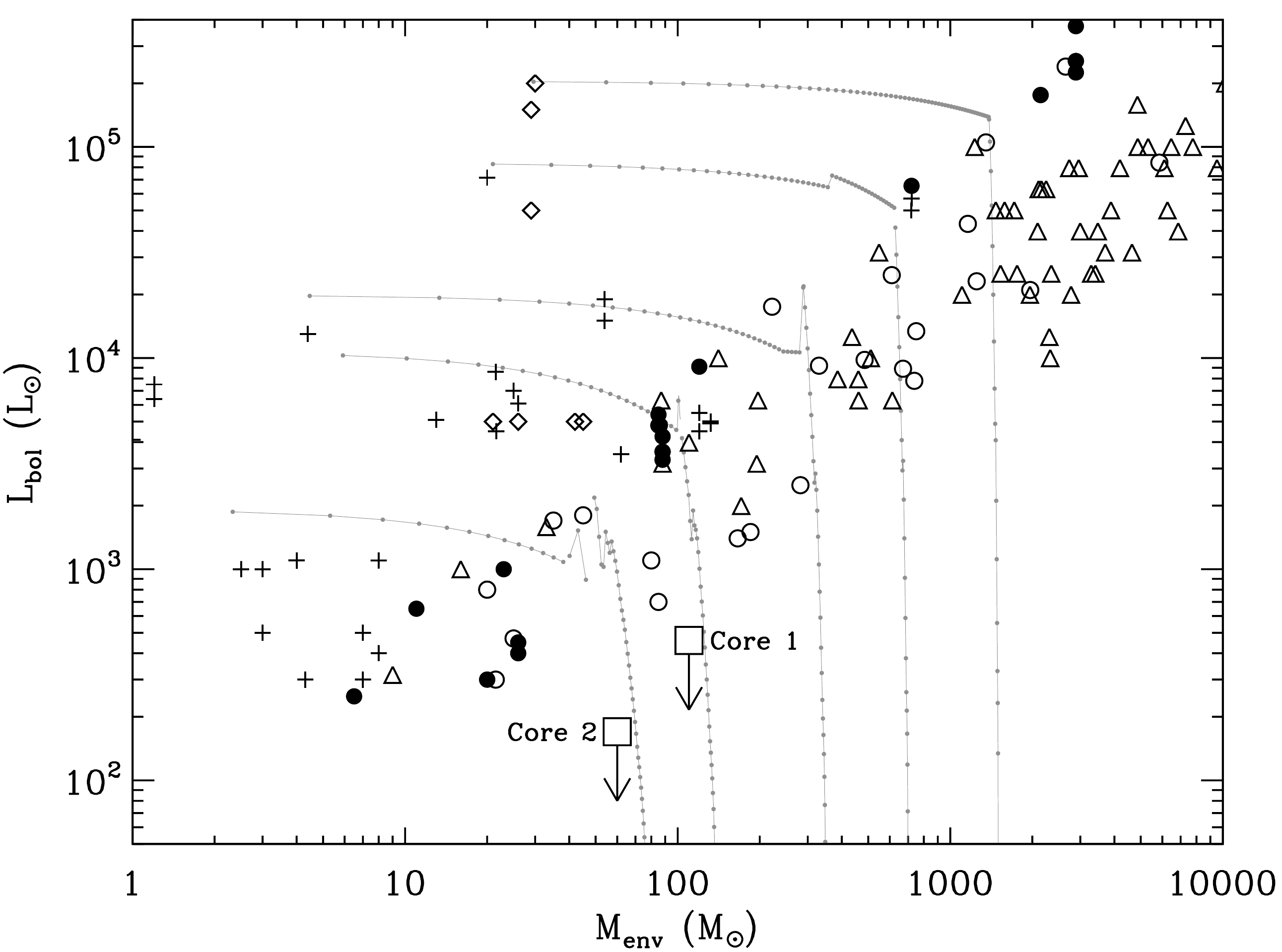}
\caption{$L_{\rm bol}$--$M_{\rm env}$ diagram displaying published
  data for massive star forming cores [open triangles
    \citep{sri02,beu02}; diamonds \citep{and08}; open circles, filled
    circles and crosses \citep[identical designations as
      in][]{mol08}]. Evolutionary tracks are from \cite{mol08} marked
  in time steps of $10^4$ years for the vertical branch and $10^5$
  years for the horizontal branch. The positions of core 1 and 2 are
  shown as upper limits in the diagram. \label{MvsL}}
\end{figure*}

The bolometric luminosity of cores versus the envelope mass, or the
\lvsm\ diagram is a useful parameter space to visualize the evolution
of cores into stars \citep[{\eg},][]{bon96}. Figure~\ref{MvsL} shows
data from several previous studies of massive star formation plotted
on the \lvsm\ diagram (see figure caption). The linear trend in
log-log space is typical for such studies. Evolutionary tracks 
based on the turbulent core model of \cite{mck03} calibrated by the
data of \cite{mol08} are also overlaid on the plot. The positions of
core 1 and 2 from this study are seen to lie well below the trend set
by previous data, another indication that they are at an early
evolutionary stage. According to the \cite{mol08} evolutionary tracks,
core 1 is on track to evolve into a star with $L_{\rm bol} \approx
10^4$\,\lsun, or of an early B spectral type while core 2 is set to 
evolve into a star with $L_{\rm bol} \approx 2 \times 10^3$\,\lsun, or
a mid-B spectral class.

The final stellar masses implied from the evolutionary tracks in
Figure~\ref{MvsL} may be underestimates. The star formation efficiency
of dense cores is expected to be between 25--50\%
\citep{mat00,alv07}. Therefore if these cores collapse monolithically
they have sufficient mass to form stars with mass between 15 and
60\,\msun. Indeed, the evolutionary models of \cite{and08} put core 1
on track to evolve into a star of $\approx 50$\,\msun\ and 
core 2 into a star with $\approx 25$\,\msun. However, it is not possible
to tell from our data whether these cores will undergo monolithic
collapse or will further fragment into smaller cores. Observations
resolving the expected fragmentation length scales ($\sim 0.03$\,pc)
are needed.

The large millimeter continuum survey of the Cygnus X region has also
revealed dense and massive cores at an early stage of evolution
\citep{mot07}. However, of the most massive cores in their sample ($M
\ge 40$\,\msun), 61\% have 21\,\micron\ {\it MSX} sources associated
with them, and the remaining either contain compact \hii, show
70\,\micron\ emission above our scaled detection limits
(S. Bontemps, F. Motte, private communication), or have SiO line
strengths and widths that indicate protostellar activity.

Therefore it seems that the cores of this study are unique, and their
existence may have implications for our understanding of
massive star formation. The selection criteria and use of
high-resolution in the sub-millimeter wave band are key components to 
these discoveries. As the characterizations of IRDCs progress
\citep[{\eg},][]{cha09} further high-resolution studies of pre-cluster
environments will provide statistics on the frequency and nature of
these kinds of objects.

\section{Summary}
This article presents interferometric mosaic observations in the
850\,\micron\ waveband toward two infrared dark clouds chosen to be
likely representations of pre-cluster environments in the Galaxy. Both
clouds are massive and dense but show no sign of massive star
formation. The most massive cores in each mosaic dominate the compact
850\,\micron\ emission and are spatially isolated near the peak of the
low-resolution continuum emission. With masses of 110 and
60\,\msun\ average densities of $\langle n \rangle \approx
10^6$\cmt\ and $\langle N \rangle \approx 1$\,g\,cm$^{-2}$ they have
the potential to form massive stars. However, upper limits to their
bolometric temperatures and luminosities, no clear indication of CO
outflow, and their relatively featureless spectra all indicate a very
early stage of evolution. Detections of ortho-\htdp$(1_{1,0} -
1_{1,1})$ and \nthp$(4-3)$ support the interpretation that these cores
are in a cold and dense state preceding significant star formation
activity. The comparison of these data with theoretical and
observational studies of massive star forming regions suggest that
these cores occupy a unique region of \lvsm\ parameter space placing
them on track to evolve into stars anywhere from mid to early B stars
up to O stars.

\acknowledgments The author is grateful for the referee's suggestions
as well as the helpful input provided by many people including David
Jewitt, John Johnson, Sylvain Bontemps, Frederique Motte, Jonathan
Williams, Emeric Le Floc'h, Qizhou Zhang, Steve Longmore, Thushara
Pillai and Jill Rathborne.

\end{document}